\documentclass[manuscript,nonacm,screen]{acmart} %
\usepackage{type1cm} 
\usepackage{graphicx} 
\usepackage{xspace} 
\usepackage{balance} 
\usepackage{booktabs} 
\usepackage{multirow} 
\usepackage{tabularx}
\usepackage{array}
\usepackage{makecell} 

\usepackage[font={bf}, tableposition=top]{caption} 
\usepackage{bold-extra} 
\usepackage{siunitx} 
\usepackage[vlined,linesnumbered,ruled,noend]{algorithm2e} 
\usepackage{microtype} 
\usepackage{xfrac} 
\usepackage{mathtools} 
\PassOptionsToPackage{hyphens}{url} 
\PassOptionsToPackage{bookmarks, pdftex, colorlinks=true, pagebackref=true, backref=page}{hyperref} 
\usepackage{cleveref} 
\PassOptionsToPackage{square,numbers}{natbib} 
\usepackage[hyperpageref]{backref} 
\usepackage{hyphenat} 
\usepackage{subcaption} 
\usepackage[export]{adjustbox} 

\renewcommand*\backref[1]{\ifx#1\relax \else (Cited on #1) \fi}

\usepackage{graphicx} 
\usepackage{longtable}
\usepackage{tabularx,ragged2e,booktabs,caption,adjustbox}
\usepackage{multicol,capt-of}
\usepackage{hhline}
\usepackage{multirow}
\usepackage{array}
\usepackage{nicematrix}\graphicspath{{img/}}
\usepackage{xcolor}
\usepackage{adjustbox}\usepackage{mathtools}
\usepackage{hhline}
\usepackage{array}\usepackage{makecell}

\newcommand{\tabhead}[1]{\textbf{#1}}
\title[]{The Role of Science in the Climate Change Discussions on Reddit}

\author{Paolo Cornale}
\affiliation{%
  \institution{ISI Foundation}
  \city{Turin}
  \country{Italy}
}
\author{Michele Tizzani}
\affiliation{%
  \institution{ISI Foundation}
  \city{Turin}
  \country{Italy}
}
\author{Fabio Ciulla}
\affiliation{%
  \institution{ISI Foundation}
  \city{Turin}
  \country{Italy}
}
\author{Kyriaki Kalimeri}
\affiliation{%
  \institution{ISI Foundation}
  \city{Turin}
  \country{Italy}
}
\author{Elisa Omodei}
\affiliation{%
  \institution{Central European University}
  \city{Vienna}
  \country{Austria}
}
\author{Daniela Paolotti}
\affiliation{%
  \institution{ISI Foundation}
  \city{Turin}
  \country{Italy}
}
\author{Yelena Mejova}
\affiliation{%
  \institution{ISI Foundation}
  \city{Turin}
  \country{Italy}
}
\email{yelenamejova@acm.org}

\begin{abstract}

Collective and individual action necessary to address climate change hinges on the public's understanding of the relevant scientific findings. 
In this study, we examine the use of scientific sources in the course of 14 years of public deliberation around climate change on one of the largest social media platforms, Reddit.
We find that only 4.0\% of the links in the Reddit posts, and 6.5\% in the comments, point to domains of scientific sources, although these rates have been increasing in the past decades. 
These links are dwarfed, however, by the citations of mass media, newspapers, and social media, the latter of which peaked especially during 2019-2020. 
Further, scientific sources are more likely to be posted by users who also post links to sources having central-left political leaning, and less so by those posting more polarized sources.
Unfortunately, scientific sources are not often used in response to links to unreliable sources.

\end{abstract}

\begin{document}

\maketitle

\pagestyle{plain}

\section{Introduction}
\label{sec:intro}

Climate change poses a critical threat that requires urgent global action. 
Despite a broad scientific agreement around a strong anthropogenic component of climate change \cite{abbass_review_2022}, 
as of 2023, only 56\% of US respondents to the Yale Climate Opinion survey thought that ``most scientists think global warming is happening''~\cite{marlon2022change,yale2023data}. 
Given the importance of public understanding of the latest scientific findings necessary for informed decision-making, in this study, we examine to what degree scientific resources are used to drive or substantiate the online discussions around climate change, in comparison to other sources, such as news and social media, including sources known to be unreliable.

Despite its privileged status in academia and industry, scientific communication competes for clicks in a cutthroat attention economy of the Web, contending with the fickle proprietary recommendation systems and shortening attention spans of their users \cite{HYLAND2023101253}.
Further, the perception, understanding, and citation of scientific literature by non-experts depend on a myriad of factors, including numerical literacy~\cite{roozenbeek2020susceptibility}, religious beliefs~\cite{schwarz2016blinded} and spirituality~\cite{browne2015going}, as well as the social context~\cite{lewandowsky2019science}.  
%
%
Social media is becoming an increasingly important source of information, with Pew Research Center concluding in 2023 that half of U.S.~adults get news at least sometimes from social media~\cite{pew2023social}.
Climate change debate has been extensively studied on Twitter~\cite{effrosynidis2022climate,karimiziarani2023toward}, yielding observations of homophilous segregation of users into like-minded camps of ``skeptics'' and ``activists'', 
which can be detected via the posted content~\cite{upadhyaya2023multi} or network analysis~\cite{williams2015network}, and which intensify during events such as the COP Climate Change conference~\cite{falkenberg2022growing}. 
However, Reddit---the fifth most visited website in the US~\cite{similarweb2024top},  which is much less studied---has been shown to display much less polarization than Twitter~\cite{de2021no} and may foster more deliberative interactions. 
The literature is lacking in the broad, longitudinal examination of scientific discourse on this platform, focusing on particular subreddits such as r/science~\cite{moriarty2024reddit,hara2019emerging}, or those relevant to the climate change debate, i.e.~r/climate or r/climateskeptics~\cite{gadanidis2020discourse,oswald2022climate,villanueva2021climate,parsa2022analyzing}. 

Further, in the US, the debate is complicated by the politicization of stances around climate change: in 2023, 23\% of Republicans considered climate change a major threat, compared to 78\% of Democrats~\cite{pew_partisan_opinion_climate_change}. 
In fact, the attitudes towards science, in general, are different among the two parties:  47\% of Republicans view science as benefiting society, compared to 69\% of Democrats~\cite{viswanathan2023americans}. 
In this study, we use 14 years of climate change-related Reddit posts and comments spanning thousands of subreddits to gauge the use of scientific resources in this deliberative space, including in the context of political interest.
Its results point to a scant, but increasing, utilization of scientific sourcing, more frequently used by the users having center-left political interests.
Alas, we find little evidence of it cited in response to information from unreliable sources.
\section{Results}

\subsection{Domain Citation and Engagement}

\begin{figure}[t]  
    \centering
    \includegraphics[width=0.90\textwidth,keepaspectratio]{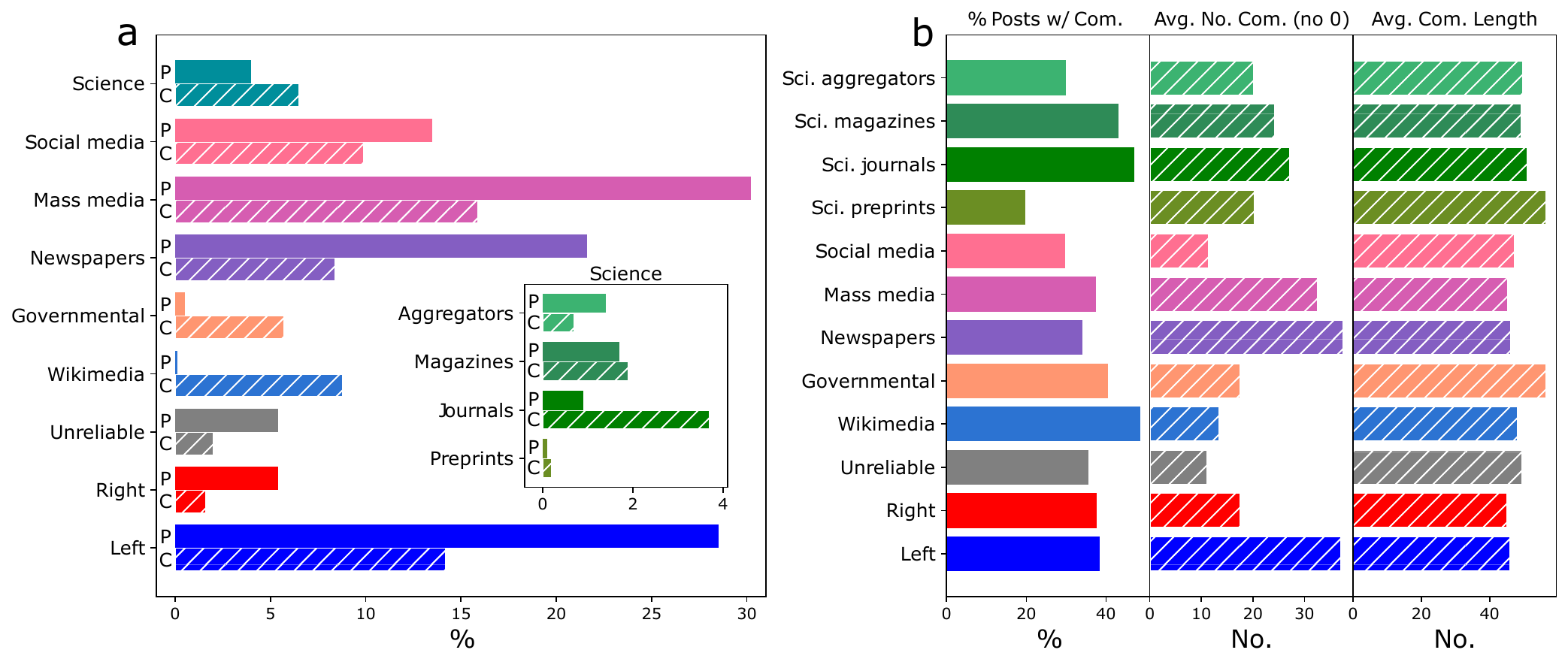}
    \includegraphics[width=0.8\textwidth,keepaspectratio]{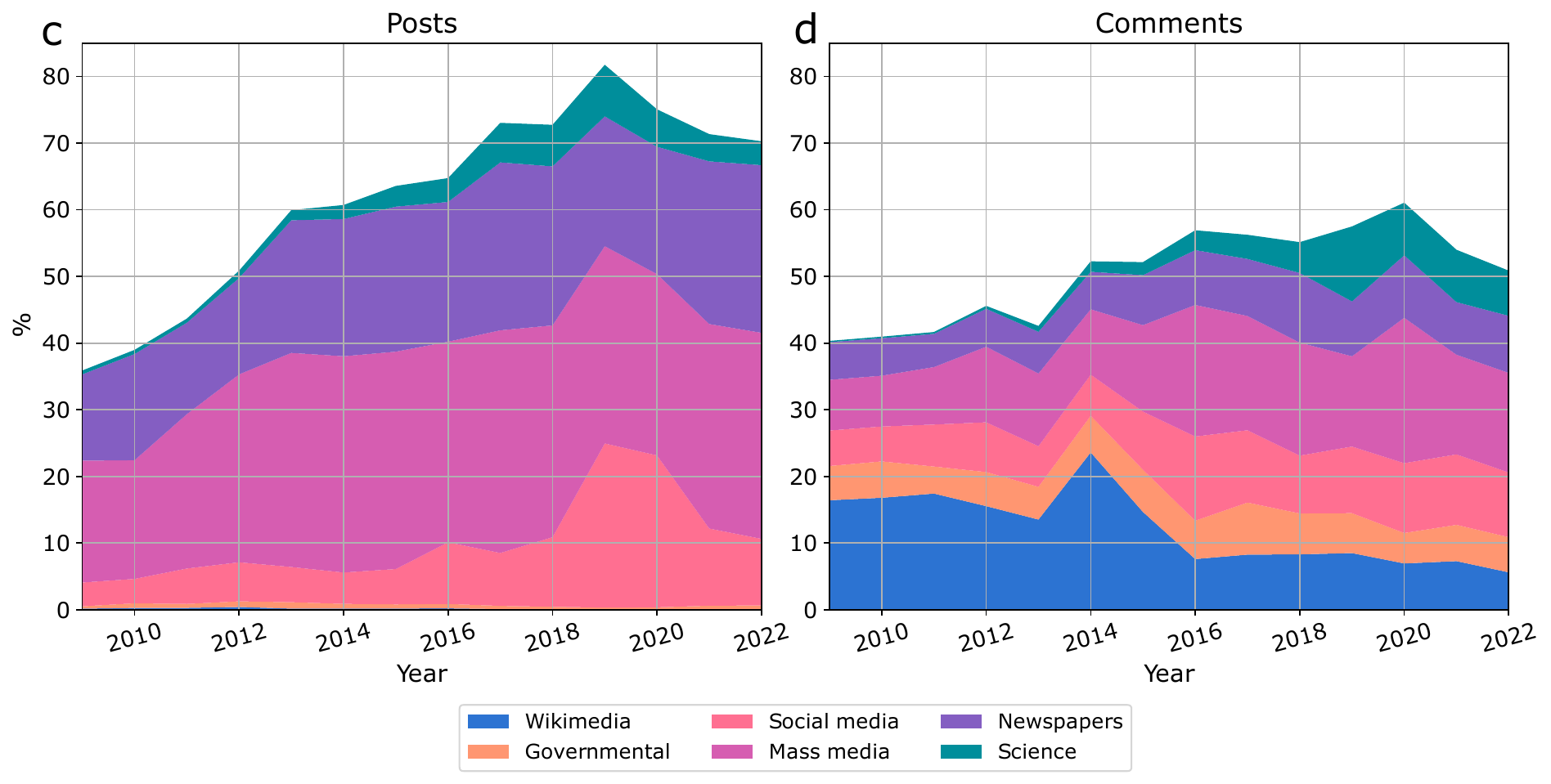}
    \caption{Statistics of URL usage in the dataset: (a) proportion of URLs in a particular category, separately for posts (preceded with ``P'') and comments (preceded with ``C'' and dashed), (b) engagement with the posts containing a URL of a particular category in terms of the percentage of posts having at least one comment, the average number of comments for posts having at least one comment, and the average length of the comment in terms of words, (c-d) proportion of URLs in a particular category in posts and comments, over time.}
    \label{fig:domain_categories_and_engagement}
\end{figure}


Using a dataset of 1.3M posts and 20.3M comments on Reddit around the topic of Climate Change, we examine the categories of URLs used in this discussion. 
We were able to categorize 69.5\% of URLs in posts and 55.2\% in comments into 6 source types (see the legend of Figure \ref{fig:domain_categories_and_engagement}a).
The largest category of URLs cited in the past 14 years is mass media (30.2\% in posts and 15.9\% in comments), followed by newspapers and social media. 
The latter (social media) has especially peaked during 2019 and 2020 (see Figures \ref{fig:domain_categories_and_engagement}c-d).
Science-related URLs appear in only 4.0\% of URLs in posts and 6.5\% in comments, though the proportion of science-related URLs has been increasing in the last decade.
These URLs to scientific domains, along with Wikimedia projects (0.1\% in posts,~8.8\% in comments) and governmental domains (0.5\%,~5.7\%) appear mostly in comments, rather than posts, pointing to their importance in the substantiating of discussion.
Among the scientific URLs, journal domains are more likely to be cited in comments, in comparison to scientific magazines and aggregators (which may have less rigorous inclusion criteria).
Interestingly, preprints, which may contain cutting-edge reporting that has not passed peer review, are the least cited scientific domain category, at 0.1\% and 0.2\% in posts and comments, respectively.
The last three statistics shown in Figure \ref{fig:domain_categories_and_engagement}a are characteristics of URLs that may overlap with those above: domains known to publish unreliable content, and those having a right or left political leaning.
Science-related domains are cited at a similar rate to domains known to be unreliable, though these are less likely to be used in the comments (5.4\% of URLs in posts and 2.0\% in comments).
Furthermore, when the political leaning of a URL is known, it is more likely to be left-leaning than right-leaning, pointing to an unequal coverage of the topic.
The quality of these URLs is also different between the two sides: 55\% of right-leaning URLs in posts and 33\% in comments are listed as unreliable, compared to only 0.01\% of left-leaning URLs in posts and 0.09\% in comments (echoing findings in \cite{robertson2023users}).



In terms of engagement, the categories of domains that receive at least one comment are those from Wikimedia, scientific journals, and magazines, followed by the governmental ones (see Figure \ref{fig:domain_categories_and_engagement}b). 
Interestingly, despite being a popular domain category, social media receives comparatively fewer comments.
However, the average length of the comments, which provides another way to measure engagement, remains remarkably stable across the domain categories, ranging around 47 - 56 words per comment. 
Whereas left-leaning domains tend to receive more comments compared to right-leaning ones, it is the domains listed as unreliable that receive the longer comments.


The distribution of the scientific links among the subreddits is extremely concentrated, with the top 10 subreddits (by the number of scientific links) accounting for 40.2\% of all science-related URLs in our dataset.
These include r/worldnews (contributing \num{20451} scientific URLs, which make up 7.2\% of all URLs in that community in our data), r/science (\num{11235}, 15.4\%), r/environment (\num{9665}, 7.2\%), r/politics (\num{8995}, 3.5\%), r/collapse (\num{7646}, 9.0\%), r/climate (\num{7140}, 9.8\%), r/climateskeptics (\num{6764}, 8.0\%), r/Futurology (\num{6422}, 8.3\%), r/climatechange (\num{5741}, 14.4\%), and r/AskReddit (\num{4600}, 8.4\%). 
Among several communities around Climate Change specifically, we also find more general ones, such as r/worldnews and r/politics, as well as r/AskReddit, attesting to the mainstream interest in the topic and the prevalence of scientific referencing even in non-specialized circles. 
Interestingly, r/climateskeptics, a community dedicated to ``Questioning climate related environmentalism'' (in the description of the subreddit), contributes over 6K links to science-related domains (which are 8.0\% of all domains on that subreddit in our dataset, a similar rate to r/worldnews and r/climate).
%
Examining the top users posting scientific links (see Supplementary section \ref{appendix_topusers}), we find several accounts (3/10) that are explicitly bots: \emph{AutoModerator}, \emph{EcoInternetNewsfeed}, and \emph{MmmBaconBot}, echoing previous findings on the prevalence of bots on the platform~\cite{ferrara2016rise}.
These bots, however, are not necessarily malicious, as they are explicit about their nature, and follow the ``bottiquette''\footnote{\url{https://www.reddit.com/wiki/bottiquette/}} \cite{hurtado2019bot,ma2020empirical}.
Given the importance of (semi-)automated accounts, we do not attempt to remove them from our dataset.


\subsection{Political Leaning and Scientific Citation}


\begin{figure}[t]  
    \centering
    \includegraphics[width=\textwidth,keepaspectratio]{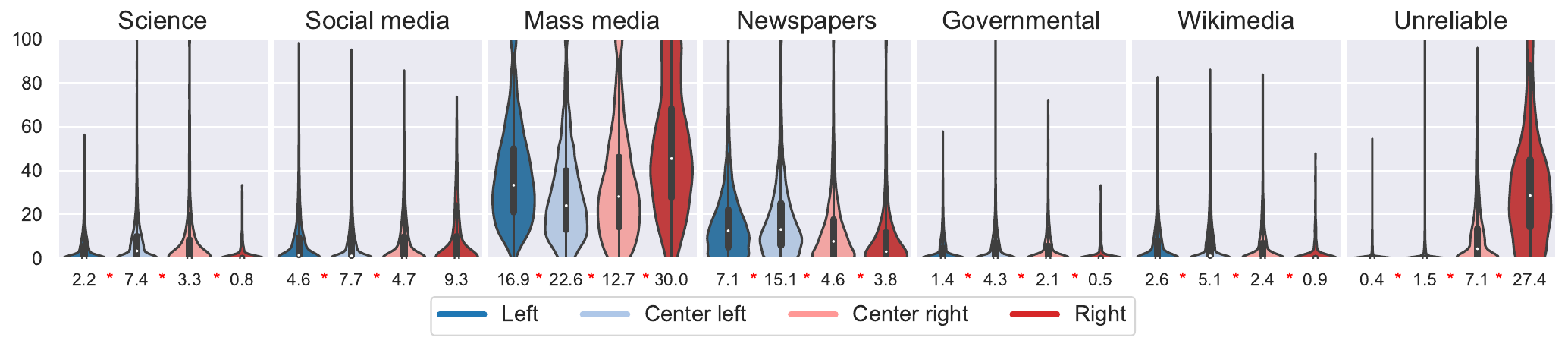} 
    \caption{Distributions of the proportion of URLs posted by a user that are from a particular category, grouped by users whose URLs have a particular political leaning. Under each distribution, the mean proportion is shown, and a * is shown between two consecutive groups if their distributions differ using the Kolmogorov-Smirnov two-sample test at $p<0.001$.
    }
    \label{fig:violin_plots_biased_users}
\end{figure}

Next, we consider users whose link posting activity puts them into the left, center-left, center-right, or right political leaning (as defined by the media bias ratings website AllSides.com).
We find that those in the center (especially center-left) are more likely to cite science than those in the extremes (see leftmost panel of Figure \ref{fig:violin_plots_biased_users}, and Supplementary Table at \ref{appendix_sharing_behavior_users}). 
In particular, out of all URLs posted by the center-left users, on average, 7.4\% are science-related, whereas 3.3\% of the ones posted by center-right users (compared to 2.2\% and 0.8\% for left and right, respectively).
Due to the large dataset sizes, we find all of the within-category differences between the consecutive groups statistically significant at $p<0.001$, despite the small effect size, except in the case of social media (which all groups of users use at roughly indistinguishable rates). 
Among the other categories, we find right-leaning users to favor mass media, whereas left, and center-left users -- newspapers. 
The most striking difference between the groups, however, is the posting of unreliable sources, which are more likely to be on the right political spectrum.


\begin{figure}[t]  
    \centering
    \includegraphics[width=0.66\textwidth,keepaspectratio]{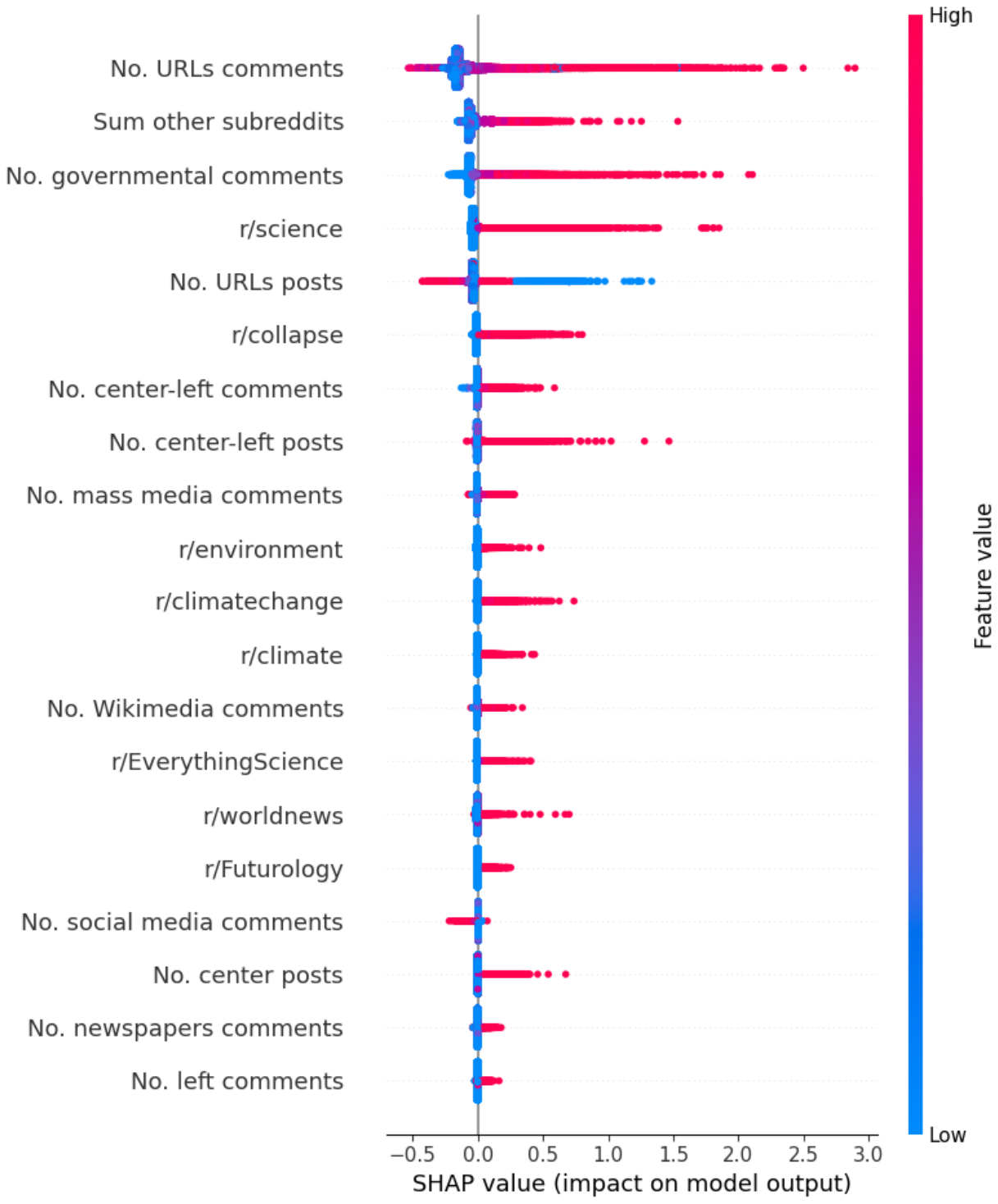}
    \caption{SHapley Additive exPlanations (SHAP), a game theoretical approach for explaining the contribution of each feature to the final output of a ML model~\citep{NIPS2017_7062}. The Random Forest model predicts how many (log-normalized) science-related URLs a user has posted in our dataset (here, ``high'' means more URLs were posted), using behavioral features including the categories of the other URLs they shared, as well as the political leaning of those URLs. Top 100 most popular subreddits are used as features, and all others are summed in ``other subreddits''. }
    \label{fig:shap}
\end{figure}

To understand the importance of political leaning in comparison with other climate change-related interests of the users, we train a model to predict how many (log-normalized) scientific URLs a user posts in our dataset (Figure \ref{fig:shap}). 
We find that posting such URLs is associated with posting on popular scientific subreddits such as r/science, r/collapse, and r/environment, and, echoing previous results, with posting links with a center-left political leaning. 
Further, posting scientific URLs is more likely by users who post many kinds of URLs in comments, 
this relationship is reversed for social media -- those posting social media links in their posts are \emph{less} likely to also post science. 


\begin{figure}[t]  
    \centering
    \includegraphics[width=0.6\textwidth,keepaspectratio]{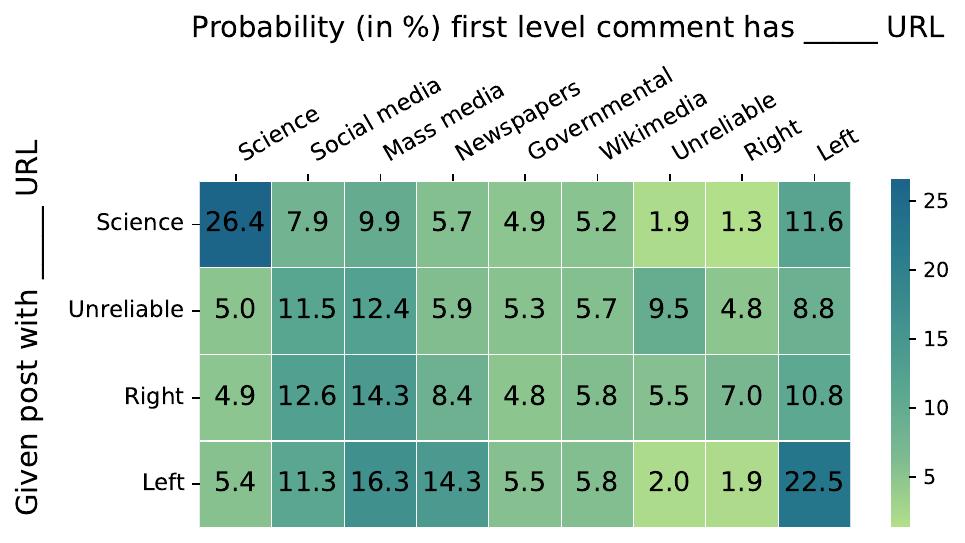}
    \caption{Conditional probability of a first-level comment with a URL containing a certain URL category (columns), given it is in response to a post having a scientific, unreliable, right- or left-leaning URL (rows).}
    \label{fig:conditional_probs}
\end{figure}

Finally, we explore the relationship between citations in posts and comments (see Figure \ref{fig:conditional_probs}). 
We find that posts with scientific URLs are much more likely to be responded to with other scientific URLs (26\% of the time). 
Unfortunately, this is not the case for the posts sharing a link to an unreliable source: only 5\% of URLs in reply to these are to science sources, instead, links from social media, mass media, and ---more likely--- other unreliable sources are posted in response.
Similarly, posts having right or left-leaning URLs have about the same chance, about 5\%, of having a scientific URL in response.

In summary, when considering individual users and the political leaning of other sources they cite, we do not find a strong polarization in terms of using scientific sources, with those on center-left being more likely to post science than on center-right. 
However, the defining feature of Reddit is the community, many of which can revolve around a topic or political theme, and which may have their own cultures beyond individual users. 
We explore these next.


\subsection{Case Study}

\begin{figure}[t]  
    \centering
    \includegraphics[width=1\textwidth,keepaspectratio]{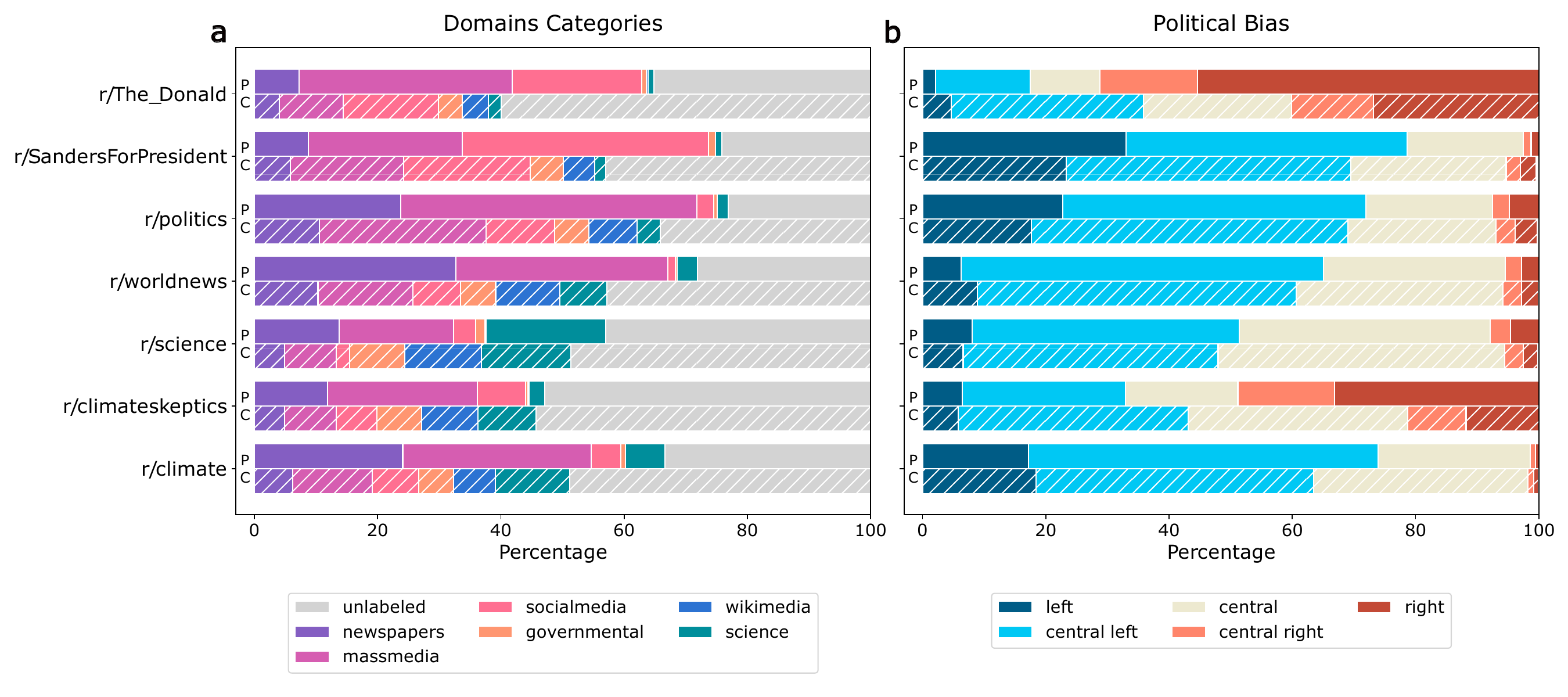}
    \caption{Case study of select subreddits, (a) the percentage of URLs having a particular domain category and (b) the percentage of URLs having a particular political leaning. Statistics are shown separately for posts (P) and comments (C).}
    \label{fig:case_study}
\end{figure}

To delve further into the nature of the Reddit communities (subreddits), we select several subreddits out of those contributing the most posts to our dataset such that they span a variety of points of view: r/climate and r/climateskeptics (largest climate-related ones); r/science, r/worldnews, and r/politics (more general ones); and r/The\_Donald (dedicated to the Republican politician Donald Trump), and r/SandersForPresident (supporting the Independent/Democrat politician Bernie Sanders). 
Note that, besides having the largest share of posts in our data out of politically-oriented subreddits, Donald Trump and Bernie Sanders represent political extremes of the right and left, respectively.
See Supplementary table \ref{appendix_casestudy} for statistics on the sizes of these subreddits.
Again, we find science, Wikipedia, and governmental links to often appear in the comments more than in posts (see Figure \ref{fig:case_study}a).
Unsurprisingly, science links are mentioned the most in the r/science subreddit, but also in the r/climate and r/worldnews.
Both of the political subreddits do not share references to science, suggesting that the political discussion of the subject is not explicitly supported by direct scientific references.
In the case of r/climateskeptics, few science URLs are included in posts, but many more are cited in the comments.
Unsurprisingly, the subreddit with the most right-leaning URLs is r/The\_Donald, with the posts having many more right-leaning URLs than comments (see Figure \ref{fig:case_study}b).
We see a similar behavior in r/climateskeptics. 
The URLs in the rest of the selected subreddits are leaning to the left, with r/SandersForPresident and r/climate ones being the most left-leaning, which is surprising, since r/climate is ostensibly not a political subreddit.
In summary, our case study suggests that, despite the Climate Change debate being highly politicized, the use of scientific evidence is lacking in the communities centered around politics, and instead is more prevalent in scientific and even in science-skeptic communities (especially in their comments).

\section{Discussion}

Thomas Jefferson is often attributed the (likely apocryphal) quote ``An educated citizenry is a vital requisite for our survival as a free people" \cite{bowman2019thomas}.
Since then, the connection between democratic deliberation and scientific education has been promoted by educators and reformers, such as John Dewey in \emph{How We Think} \cite{dewey1933how}, and more recently by the U.S.~National Research Council, positing that ``knowledge of science and engineering is required to
engage with the major public policy issues of today'' \cite{nrc2012framework}. 
Our findings show that, in the Reddit discussions of climate change, scientific sources have been dwarfed by links to news and social media, although the share of links to scientific resources has increased in the past decade. 
When they do appear, they are more likely to be in the comments (along with the links to governmental sources and Wikipedia), pointing to the importance of these resources to the deliberative process around this topic.
Unfortunately, we find that scientific links are much more likely to be posted in response to posts with other scientific links, whereas posts having links to unreliable sources do not often receive scientific links in their replies.
Instead, other unreliable sources are more likely to be cited.

Meanwhile, surveys show that between 2009 and 2019 (roughly in the duration of the examined data), the share of US respondents who acknowledge an increase in average global temperature rose by 8 percentage points, and the share who believe that humans have contributed to this rose by 11 percentage points \cite{carlsson2021climate}.
Whether the use of scientific resources contributed to this change of opinion is questionable. 
Experimental results suggest articles linking to scientific articles promote greater trust, however linking to any mainstream media may have the same effect \cite{verma2017human}. 
We find that scientific links appear not only in communities asserting the existence of anthropogenic climate change, but also in those ``skeptical'' of it.
According to the 2024 report by the Center for Countering Digital Health, climate change denialism has been morphing in the past few years, as global temperatures rose dramatically \cite{ccdg2024new}. 
Instead of opposing the concept of climate change itself, the ``New Denial'' themes include ``the impacts of global warming are beneficial or harmless'', ``climate solutions won't work'' and ``climate science and the climate movement are unreliable''.
The report points specifically to social media companies (including Instagram, Facebook, TikTok, and X (previously Twitter)) as potentially benefiting from the popularity of such content, and allowing for the monetization and direct profit for its creators.
What role the latest scientific evidence may play in tackling this New Denial themes is an open question.

However, the mere presence of scientific evidence may not necessarily correspond to a cross-partisan conversation.
Psychology literature suggests that individuals more knowledgeable on an issue are more susceptible to selection bias and motivated reasoning \cite{bushey2012everyday,hannon2022knowledgeable,kahan2013ideology}.
Our discovery that 9.4\% of the URLs in r/climateskeptics community's comments are scientific links points to the active use of science in the community. 
Previous studies found that, for instance, vaccine skeptics often express respect for the scientific method and are interested in the rigorous scientific examination of matters affecting them personally \cite{koltai2017questioning}. 
Further, the level of one's education may affect the trust in climate science via the perception of having a lower or higher social status \cite{hoekstra2024educational}. 
To some, attitudes towards climate science may be ``not just an opinion on an issue, but [an] aspect of self that defines who they are, what they stand for, and who they stand with (and against)'' \cite{bliuc2015public}. 
Thus, the citation of science may be a part of the construction of self-evaluation as ``eco-habitus'', a concept favoring environmental actions and engagement in a ``green'' lifestyle \cite{kennedy2019eco}. 
\citet{hoekstra2024educational} found that such subjective view of one's social status may contribute to the distrust in climate science.
What role the citation of scientific literature plays in the individuals' formation of a self-image is an interesting future research direction.

In the U.S., a major aspect of such self-image may be one's political affiliation.
The stated policies of the two parties concerning climate change differ substantially: whereas the Democratic party elites have been consistently supportive of the climate consensus~\cite{merkley2021party}, the Republican party, and especially its neoliberal champions, argue that environmentalists in the government ``intrude'' on society by curtailing consumer choice and property rights~\cite{antonio2011unbearable,smith2024polarisation}.
Furthermore, concerning science and academia, in the past decade, there has been a sharp decline among Republicans of those who ``believe that colleges and universities have a ``positive effect'' on the country''~\cite{atkeson2019partisan}.
However, in our case study, we find that both the Republican (r/The\_Donald) and Democrat/Independent (r/SandersForPresident) community had a negligible number of links to scientific sources. 
The little scientific citation that does circulate in politically-oriented discussions may be influenced by the communiqu\'{e}s of NGOs, think tanks, and government reports (papers cited by such reports are more likely to be highly cited \cite{bornmann2022relevant}), each bringing its own agenda.
Conversely, the perception of a scientific source may be affected by the political stances of its editors \cite{zhang2023political}.
Instead, social media dominates these communities' links (21\% in posts on r/The\_Donald and 40\% on r/SandersForPresident).
As the major social media platforms have been thoroughly documented for spreading scientific misinformation \cite{osman2022youtube,do2022infodemics}, and some smaller ones boasting even more permissive policies \cite{bar2023new}, the extensive use of these as a resource for policy and science discussions is highly concerning and should be further investigated.

\section{Methodology}

\subsection{Reddit Dataset Collection}

We choose a high-precision keyword-based approach to collect a dataset related to Climate Change. 
This method has been used extensively in the literature \cite{treen2022discussion,shah2021climate,parsa2022analyzing}, as it has been shown that conversation relevant to climate change happens in many subreddits, most of them not devoted exclusively to this topic. 
We use the data collected by Pushshift via the Reddit API \cite{baumgartner2020pushshift} in the 168 months between January 2009 and December 2022. 
Using manual examination of the dataset, we compose a set of 64 bigrams that maximize the coverage and minimize false positives (see Supplementary \ref{appendix_keywords}). 
We first collect the posts and comments that contain at least one of these bigrams in the title (for posts) or the content (posts and comments). 
We then add to our dataset all comments to the selected posts regardless of their match to the keywords list.
The matching resulted in $1,301,970$ posts and $6,428,051$ comments, and an additional $15,273,754$ comments in response to the posts.
After removing duplicates, a total of $20,279,912$ comments remain in our dataset.
The volume of posting increases over time, and peaks around 2019-2020 (see Supplementary graph in \ref{appendix_volume}).


To assess the relevance of the resulting dataset, all six authors, who are fluent in English, manually labeled a random sample of the submissions and comments that matched keywords by using three labels: "relevant", "partially relevant" and "non-relevant". 
After labeling 324 posts, we find that 84.9 \% were labeled as relevant, 10.5\% as partially relevant, and 4.6\% as non-relevant. 
The Cohen's kappa, computed on a sample of 60 posts, is $\kappa=0.55$.
After labeling 384 comments that matched keywords, 79\% were judged to be relevant, 16\% as partially relevant, and 5\% as non-relevant.
The Cohen's kappa, computed on a sample of 66 comments, is $\kappa=0.50$.
The comments in response to relevant posts were often too short and uninformative to be accurately labeled. We assume they are relevant in the discussion, because they are answers to posts mostly considered pertinent.



\subsection{Sources of Information}
\label{sec:infosources}

As the focus of this study is the citation of different kinds of information, and specifically science, we consider the URLs shared in the posts and comments that talk about climate change. 
We disregard URLs pointing to Reddit itself and resolve URLs to web.archive.org or archive.is --the two most popular archive services for web pages in our dataset.  
In our study, we consider only the subreddits that have shared at least 10 URLs in the 14 years of the dataset. 
The remaining 7,837 subreddits (12.19\% of the total) allow us to keep 778,728 (96.83\%) URLs shared in the posts, and 2,929,061 (99.00\%) URLs of the comments.

We then focus on the domains of the extracted URLs and define six categories characterizing them as a source of information: social media, newspapers, mass media, WikiMedia, governmental sources, and scientific sources (see Supplementary \ref{appendix_categories} for a summary of the categories).
We use both outside sources to create the lists of domains of interest, as well as examine the top 100 domains used in our dataset that are not a part of any list. 
By social media, we consider the six most popular ones in our dataset: Twitter, YouTube, Facebook, Instagram, LinkedIn, and Discord\footnote{`twitter.com', `youtube.com', `facebook.com', `instagram.com', `linkedin.com', `discord.gg', `discord.com'}.
We obtain the list of 4,898 newspapers (with their domains) from the media portal \emph{Scimago}\footnote{\url{https://www.scimagomedia.com/rankings.php}}, with the addition of Financial Times "ft.com" from the manual domain examination. 
The list of mass media is taken from the media ranking website AllSides\footnote{https://www.allsides.com/media-bias/ratings}, in particular by looking at four types of sources: News Media, Reference, Fact Check, and Think Tank/Policy Groups. We do not consider the individual authors. After removing domains already in other categories, the mass media list has 1,623 domains. We cleaned this list by removing the sources added to other categories, and added 9 domains from manual examination\footnote{`huffingtonpost.com', `businessinsider.com', `abc.net.au', `pbs.twimg.com',  `msn.com', `news.gallup.com', `nationalobserver.com', `ctvnews.ca', `oann.com'}. 
The WikiMedia category contains all the domains from the official webpage of the Wikimedia Foundation Project\footnote{https://wikimediafoundation.org/our-work/wikimedia-projects/\#a2-collections https://foundation.wikimedia.org/wiki/Home}, adding also ``upload.wikimedia.org", used for access to the media files. In total, there is an amount of 28 domains in this category. 
For the governmental sources, we consider the domains ending with ``.gov'', dropping the 4 that appear in the scientific journal list (`cdc.gov', `ehp.niehs.nih.gov', `nist.gov', `wwwnc.cdc.gov') and `eric.ed.gov' that is a preprint domain. In total, this category has 3,194 .gov domains. To this ones, we added 39 domains collected from the official UN website\footnote{https://www.un.org/en/about-us/un-system}. 

For the scientific sources, we make a distinction between four subcategories: magazines, journals, scientific news aggregators, and preprints. 

\begin{itemize}

    \item Magazines -- the mainstream scientific source of information written for the general non-expert public. We obtain the list of the most popular English-language magazines, with their websites, from Wikipedia\footnote{\url{https://en.wikipedia.org/wiki/Category:Science_and_technology_magazines_by_country}} and manual research on the Web. After removing the few peer-reviewed ones (because they are considered journals), we collected 185 magazine URLs.

    \item Journals -- a peer-reviewed publication, written by and for experts. We obtain a list of journals by scraping the platform \emph{Web of Science}\footnote{\url{https://wosjournal.com/}}. After removing the journals without a URL (generally, they have only the link to the publisher, which could be misleading), we manually added different variations of domains, resulting in 1943 different domains.

    \item Scientific news aggregators -- web applications that aggregate scientific or technological content from different sources, which are not necessarily peer-reviewed. After manual research on the Web, we collect 5: ``sciencedaily.com", ``phys.org", ``eurekalert.org", ``esciencenews.com" and ``researchgate.net". 

    \item Preprints -- scientific papers published before the peer-review process. We scrape their directories from the \emph{Directory of Open Access Preprint Repositories} webpage\footnote{\url{https://doapr.coar-repositories.org/repositories/}}, collecting 83 preprint domains.
\end{itemize}

We make the full list of domains and their categories available to the research community\footnote{\url{https://anonymous.4open.science/r/domains_types-31D5/README.md}}. \\

To supplement our understanding of the quality of the domains, we use previous literature and reputable sources to create a list of unreliable domains.
For this purpose, we use the Wikipedia Lists of fake news websites\footnote{\url{https://en.wikipedia.org/wiki/List\_of\_fake\_news\_websites\#Lists}\\ \url{https://en.wikipedia.org/wiki/List\_of\_miscellaneous\_fake\_news\_websites}}, Media Bias Fact-Check (MBFC) lists of conspiracy and fake news webistes\footnote{\url{https://mediabiasfactcheck.com/conspiracy/}\\  \url{https://mediabiasfactcheck.com/fake-news/}}
and previous literature on climate change on Reddit \cite{gadanidis2020discourse}. 
Additionally, we use the media ranking website All Sides political bias domain labels, merging ``right'' and ``center right'' into ``right'' labels and similarly for ``left".

Finally, we enrich the domain list with (US-centric) political leaning information from Allsides that provides five labels: left, left-center, center, right-center, and right. 
For computing statistics, we merge ``right'' and ``center right'' into ``right'' labels and similarly for ``left".
On the other hand, when we compute a political bias score for each subreddit based on the URLs appearing in its posts and comments, we assign numerical values to these labels from -2 to 2 (from ``left'' to ``right'') and average these scores for each subreddit. 
We perform the same computation for the users to summarize the political leaning of the URLs they have shared in our dataset. 
To avoid noise due to sparsity (wherein not enough URLs were posted by each user), we examine the distribution of these scores, compute the Jansen-Shannon distance between the consecutive distributions, and determine the cutoff of 5 URLs with bias labels, such that the distribution is stable (see Supplementary \ref{appendix_userbias}).

\subsection{Modeling Scientific URL Use}

We compute the political bias score for the users in a similar way we have for the subreddits and we keep only the users that shared at least 5 biased links (see appendix \ref{appendix_userbias} for details). Therefore, we consider 26,620 users. 
In order to find the most relevant attributes related to the sharing of scientific links, we decided to build an explanatory model by focusing on the different number of categories of domains and on the top 100 subreddits (by the number of links) in our dataset. The remaining subreddits are placed in the "other subreddits" variable. We remove every scientific reference in the design matrix, both in the politically biased links (some scientific sources have a "central" bias) and in the number of URLs shared on the subreddits.
After shuffling the data, we take the logarithm of each of these numeric values (to which 1 was previously added~\cite{pither2023introduction}, to compute the logarithm of the zeros) because the data is highly skewed/asymmetrical. and run a Random Forest model having the number of scientific links as the dependent variable. We use 3-fold cross-validation to find the best values of the hyperparameters (number of trees and their depth) obtaining an average score (i.e. mean accuracy) of 0.63. Finally, we explain the model with SHAP\footnote{\url{https://shap.readthedocs.io/en/latest/index.html}} (SHapley Additive exPlanations) a method that uses the Shapley values from cooperative game theory to explain how the coefficients of the model interact with the output. 

\subsection{Conditional Probability of URL in Response}

To better understand how the different URL categories are used in response to potentially politically biased content, we compute the conditional probabilities as follows.
Given a post with a URL of a particular category, we compute the conditional probability that a URL of another category is used in a first-level comment to that post. 
Note that, for this computation, we consider all posts that have at least one URL, and all first-level comments to them that have at least one URL.

\bibliographystyle{ACM-Reference-Format}
\bibliography{referencesM,references}

\appendix

\pagebreak
\section{Supplementary Material}
\label{supplementary}

\subsection{Keywords Regarding Climate Change}
\label{appendix_keywords}

"climate change", "global warming",  "climate crisis", "climate action", "climate strike", "carbon tax", "climate accord", "climate activist", "climate agenda", "climate agreement", "climate alarmism", "climategate", "climate anxiety", "changing climate", "climate camp", "climate catastrophe", "climate chaos", "climate commitment", "climate concern", "climate conference", "climate consensus", "climate cooperation",  "climate council", "climate crises", "climate deal", "climate debate", "climate denial", "climate denier", "climate disaster", "climate dissent", "climate emergency", "climate fight", "climate goal", "climate impact", "climate issue", "climate justice", "climate march", "climate meeting", "climate migrant", "climate model", "climate movement", "climate myth", "climate pact", "climate panel", "climate plan", "climate policies", "climate policy", "climate-change", "climate protection", "climate protest", "climate rebellion", "global heating", "climate report", "climate research", "climate risk", "climate scam", "climate sceptic", "climate science", "climate scientist", "climate skeptic", "climate summit", "climate talk", "climate threat", "climatechange"

\subsection{URL domain categories and their sources}
\label{appendix_categories}

\begin{table}[H]
  \centering
  \renewcommand{\arraystretch}{1.2}
  \begin{NiceTabular}{|p{5cm}|c|c|}
    \hline
\textbf{Domain category}  & \textbf{No. domains} & \textbf{Sources}\\ \hline\hline   
Social Media & 7 & Extracted from data
\\ \hline 
Wikimedia Foundation Project & 28 & Wikimedia Project 
\\ \hline 
Newspapers & 4898 & Scimago Media Rankings 
\\ \hline 
Mass Media & 1621 & AllSides and Extracted from data 
\\ \hline 
Governmental: \emph{.gov} domains & 3194 & Extracted from data
\\ \hline 
Governmental: United Nations System & 39 &  United Nations Webpage
\\ \hline 
Scientific Aggregators & 5 & Wikipedia and extracted from data
\\ \hline 
Scientific Magazines & 185 & Wikipedia
\\ \hline 
Scientific Journals & 1943 & Web of Science 
\\ \hline 
Scientific Preprints & 83 &  Directory of Open Access Preprint Repositories 
\\ \hline\hline
Unreliable & 2469  & Wikipedia, Media Bias/Fact Check
\\ \hline 
Right political leaning &  159 & AllSides
\\ \hline 
Left political leaning & 322  & AllSides
\\ \hline 

\end{NiceTabular}
\caption{Summary of the domain categories and their sources. Those above double-line are mutually exclusive, whereas those below (unreliable and political leanings) are not.} 
\label{tab:summary_labels}
\end{table}

\subsection{Volume of Reddit Posts and Comments around Climate Change}
\label{appendix_volume}

\begin{figure}[t]  
    \centering
    \includegraphics[width=0.9\textwidth,keepaspectratio]{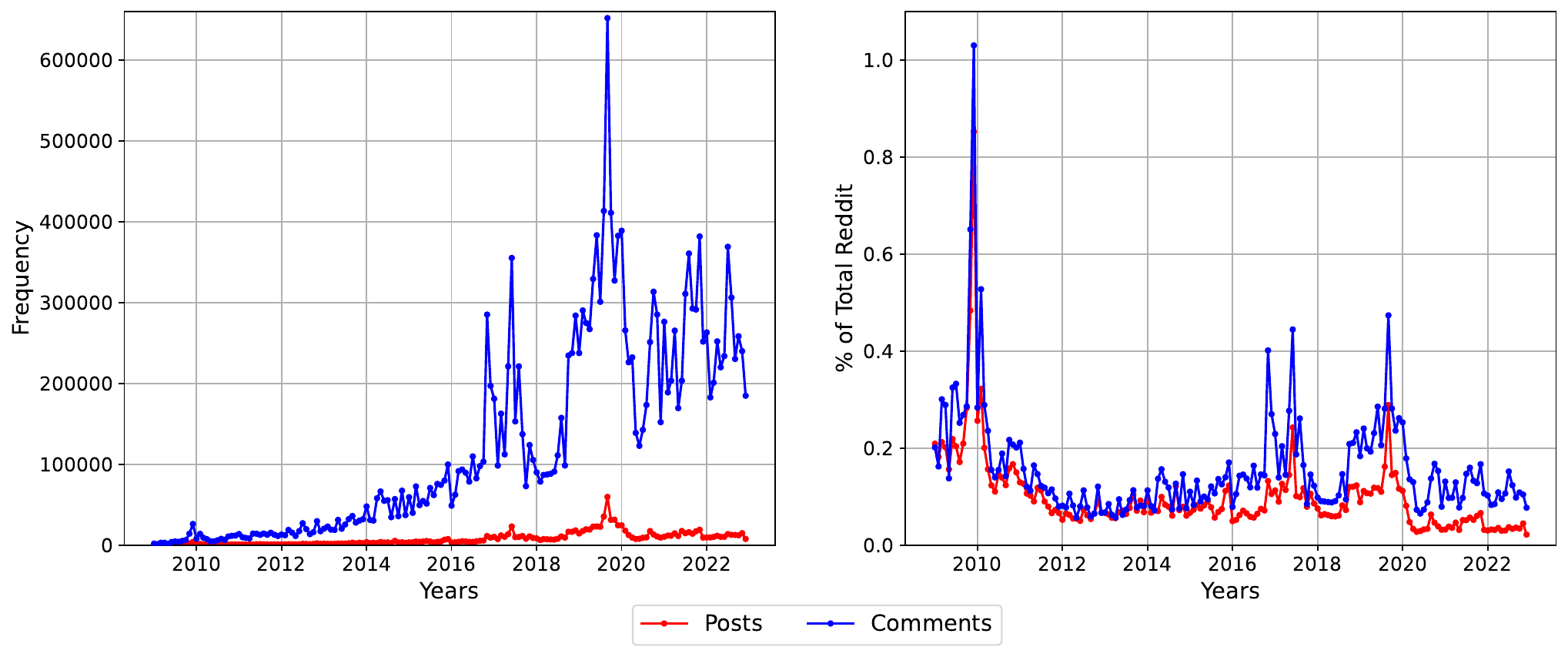}
    \caption{Temporal statistics of the dataset (both posts and comments), aggregated by month.}
    \label{fig:plots_dataset}
\end{figure}

Figures \ref{fig:plots_dataset} (a, b) show the temporal statistics of the dataset (both posts and comments), aggregated by month. 
Figure \ref{fig:plots_dataset} (a) shows the raw number of posts and comments, and (b) shows the proportion of these posts and comments of the whole Reddit posting volume.
Although we find that the volume of posts and comments increases over time (peaking especially in 2019-2020), as a share of the overall Reddit posting volume, the years 2009, 2017, and 2019 show an especially higher focus on the topic of the overall Reddit activity.

\subsection{Top Users Posting Scientific Links}
\label{appendix_topusers}

Turning to users who have contributed the most number of scientific links, at the top we find \emph{ILikeNeurons} (\num{15685}), 
\emph{worldnews} (\num{20451}), 
\emph{BurnerAcc2020} (\num{3254}), 
\emph{AutoModerator} (\num{2962}), 
\emph{ZephirAWT} (\num{2926}), 
\emph{Jaagsiekte} (\num{2412}), 
\emph{EcoInternetNewsfeed} (\num{3382}), 
\emph{MmmBaconBot} (\num{3382}), 
\emph{kamjaxx} (\num{1093}), and
\emph{avogadros\_number} (\num{1684}).
These top 10 users contribute 16.6\% of all scientific URLs in our data.

\subsection{Link-based Political Leaning of Users}
\label{appendix_userbias}

\begin{figure}[H]  
    \centering
    \includegraphics[width=0.86\textwidth,keepaspectratio]{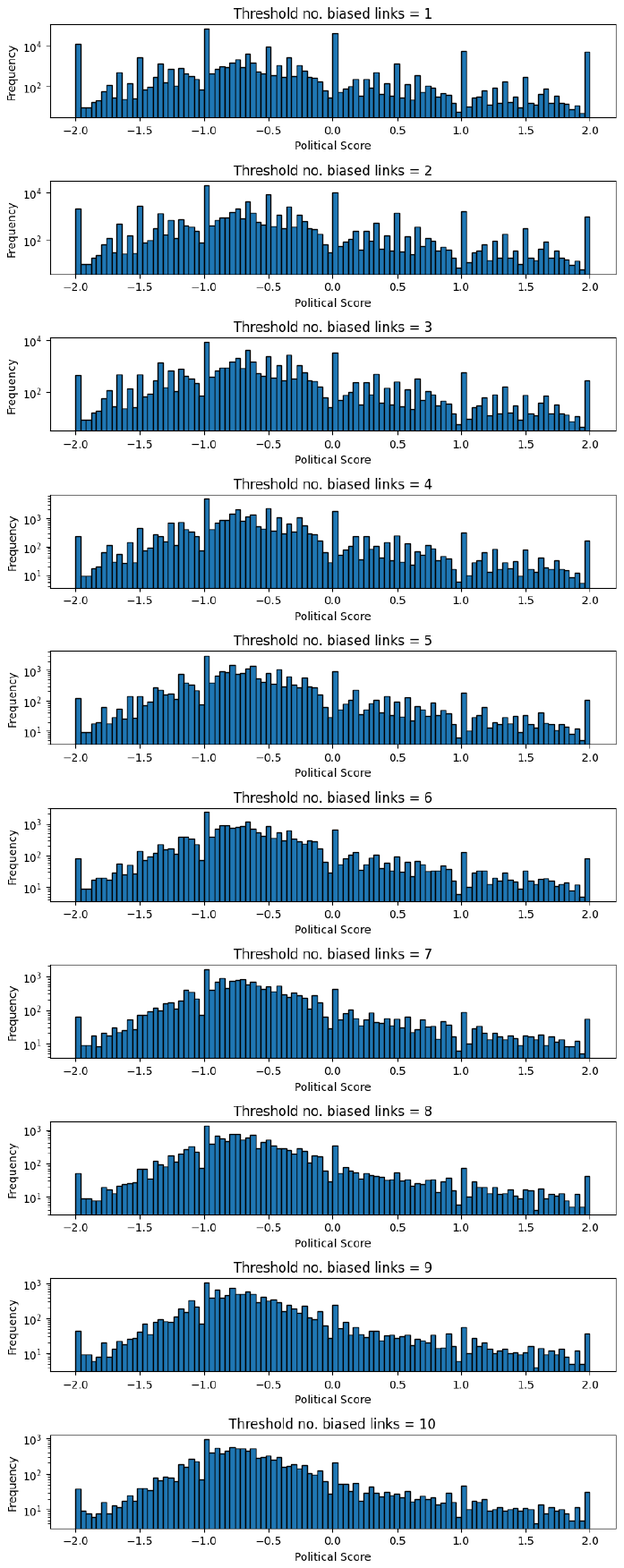}
    \caption{\textbf{Distributions of the users' political score with different thresholds of the number of biased links shared}}
    \label{fig:threshold_biased_users}
\end{figure}

We use information about the domains to which each user has linked to in their posts and comments as indicators of the sources of information they favor to compute a user-specific political leaning score.
To each link we assign a numerical political bias score  mapped to AllSides from -2 to 2 (from ``left'' to ``right'').
We then average the scores of all links a user has posted.
To gauge the sparsity of the dataset, we plot the distribution of the user scores in Figure \ref{fig:threshold_biased_users} and compute the Jensen-Shannon Distance between couples of consecutive distributions. 
We choose as the threshold 5, as the first time the Jensen-Shannon Distance between the distributions is lower than 0.2 is between thresholds 5 and 6. 
Thus, for further analysis we consider only the users that shared at least 5 URLs with a known bias score, resulting in \num{24453}, which is 14.6\% of the users in the whole dataset.

\subsection{URL Sharing Behavior of Users with Political Leaning}
\label{appendix_sharing_behavior_users}

\begin{table}[h]
\caption{Mean, median, and standard deviation of the proportion of URLs of a certain category posted by users having a specific political leaning (identified via other URLs they have shared).}
\label{tab:sharing_behavior_users}
\centering
\begin{NiceTabular}{|c|ccc|ccc|ccc|ccc|}
\hline

\tabhead{} & \multicolumn{3}{c|}{\tabhead{Left}} & \multicolumn{3}{c|}{\tabhead{Center-left}} & \multicolumn{3}{c|}{\tabhead{Center-right}} & \multicolumn{3}{c|}{\tabhead{Right}} \\
\cmidrule(lr){2-4} \cmidrule(lr){5-7} \cmidrule(lr){8-10} \cmidrule(lr){11-13}
\tabhead{} & \textbf{Mean} & \textbf{Med.} & \textbf{Std} & \textbf{Mean} & \textbf{Med.} & \textbf{Std} & \textbf{Mean} & \textbf{Med.} & \textbf{Std} & \textbf{Mean} & \textbf{Med.} & \textbf{Std} \\
\hline\hline 
Science &  2.24 & 0 & 11.31 &
7.45 & 1 & 133.36 &
3.35 & 0 & 12.82 & 
0.80 & 0 & 5.02 \\
Social media &  4.59 & 1 & 30.38 &
7.72 & 1 & 192.53  &
4.76 & 0 & 35.17 &
9.34 & 0 & 103.54 \\
Mass media &  16.90 & 6 & 74.50 & 
22.63 & 6 & 316.76 &
12.72 & 6 & 39.23 &
30.04 & 7 & 145.37\\
Newspapers & 7.06 & 3 & 35.08 & 
15.06 & 3 & 265.84 &
4.61 & 2 & 13.72 & 
3.79 & 1 & 14.85   \\
Governmental &  1.45 & 0 & 6.00 & 
4.35 & 1 &  65.97 &
2.07 & 0 & 8.03 &
0.53 & 0 & 1.95 \\
Wikimedia & 2.57 & 0 & 9.42 & 
5.10 & 1 & 69.13 &
2.38 & 0 & 8.30 &
0.92 & 0 & 4.95 \\
Unreliable &  0.43 & 0 & 4.84 & 
1.56 & 0 & 33.50 &
7.12 & 1 & 51.35 &
27.41 & 5 & 132.50\\
\hline
\end{NiceTabular}
\end{table}

\subsection{Subreddits in Case Study}
\label{appendix_casestudy}

We chose a selection of subreddits to explore in more depth out of those which have contributed the most posts to our dataset. 
For each subreddit, the table below shows its description (from Reddit), its membership in terms of users, number of posts and comments in our dataset, and the number of URLs present in these.
All statistics were collected in April 2024, except those for The\_Donald, which was banned at the time, for which its Wikipedia page was used\footnote{https://en.wikipedia.org/wiki/R/The\_Donald}.

\begin{table}[H]
\caption{Statistics on subreddits considered in the case study.}
\label{tab:special_subs}
\centering
\begin{NiceTabular}{|p{7cm}|c|c|c|}
\hline 
\tabhead{Subreddit, Description} & \tabhead{Members} & \tabhead{No. Posts \& Comments} & \tabhead{No. URLs} \\
\midrule \hline 

\textbf{r/climate}: Information about the world's climate. Truthful and accurate information about the world's climate, as well as related activism and politics & \num{187919} & \num{175833} & \num{73047}  \\ \hline 
\textbf{r/climateskeptics}: Climate Skeptics: Trying to see through the alarmism. Questioning climate related environmentalism & \num{43440} & \num{281120} & \num{84897}  \\ \hline 
\textbf{r/science}: This community is a place to share and discuss new scientific research. Read about the latest advances in astronomy, biology, medicine, physics, social science, and more. Find and submit new publications and popular science coverage of current research & 32M & \num{591697} & \num{72873}  \\ \hline 
\textbf{r/worldnews}: place for major news from around the world, excluding US-internal news & 36M & \num{2123981} & \num{282328}  \\ \hline 
\textbf{r/politics}: is for news and discussion about U.S. politics & 8.5M & \num{633062} & \num{256341}  \\ \hline 
\textbf{r/The\_Donald}: is a never-ending rally dedicated to the 45th President of the United States, Donald J. Trump  & max \num{790000}  & \num{212769} & \num{38160}  \\ \hline 
\textbf{r/SandersForPresident}: Bernie Sanders 2024 & \num{506918} & \num{281120} & \num{84897} \\ \hline 
\end{NiceTabular}
\end{table}

\end{document}